# Do Rainbow Trout and Their Hybrids Outcompete Cutthroat Trout in a Lentic Ecosystem?

Joshua M. Courtney,[1] Amy C. Courtney,[1] and Michael W. Courtney[2*]

1 BTG Research, P.O. Box 62541, Colorado Springs, Colorado, United States of America,
2 United States Air Force Academy,[1] 2354 Fairchild Drive, USAF Academy, Colorado, United States of America
*E-mail: Michael_Courtney@alum.mit.edu

**Abstract**
Much has been written about introduced rainbow trout interbreeding and outcompeting native cutthroat trout. However, the specific mechanisms by which rainbow trout and their hybrids outcompete cutthroat trout have not been thoroughly explored, and most of the published data is limited to lotic ecosystems.  Samples of Snake River cutthroat trout (*Oncorhynchus clarkii bouvieri*), the rainbow-cutthroat hybrid, the cutbow trout (*Onchorhynchus mykiss x clarkii*), and rainbow trout (*Oncorhynchus mykiss*), were obtained from a lentic ecosystem (Eleven Mile Reservoir, Colorado) by creel surveys conducted from May to October, 2012.  The total length and weight of each fish was measured and the relative condition factor of each fish was computed using expected weight from weight-length relationships from the Colorado Division of Parks and Wildlife (CDPW).  Data from the CDPW collected from 2003 – 2010 in the same lentic ecosystem were used to compute relative condition factors for additional comparison, as was independent creel survey data from 2011.  The data was also compared with minimum, $25^{th}$ percentile, mean, $75^{th}$ percentile, and maximum weight-length curves generated from independent North American data.  Cutthroat trout were plump: the mean relative condition factor of the cutthroat trout was 112.0% (± 1.0%).  Cutbow hybrid trout were close to the expected weights with a mean relative condition factor of 99.8% (± 0.6%).  Rainbow trout were thinner with a mean relative condition factor of 96.4% (± 1.4%).  Comparing mean relative condition factors of CDPW data from earlier years and plotting the 2012 data relative to percentile curves also shows the same trend of cutthroat trout being plumper than expected and rainbow trout being thinner than the cutthroat trout, with the hybrid cutbow trout in between.  This data supports the hypothesis that rainbow trout do not outcompete cutthroat trout in lentic ecosystems.  Comparison with data from three other Colorado reservoirs also shows that cutthroat trout tend to be more plump than rainbow trout and their hybrids in sympatric lentic ecosystems.

---

## 1. Introduction

Since the late 1800s, most taxa of cutthroat trout (*Oncorhynchus clarkii*) have experienced dramatic reductions in abundance and distribution, with the greenback cutthroat trout (*O. clarkii stomias*) being listed as threatened under the U.S. Endangered Species Act [1, 2]. Many authors cite introduced rainbow trout (*Oncorhynchus mykiss*) as having a great impact on native cutthroat trout through hybridization and competition [2-5]. Most studies have focused on competition in lotic ecosystems (rivers and streams) [6-8]. It may seem reasonable that similar relationships hold for lentic ecosystems (lakes) also, but little work has been published on competition between cutthroat trout and rainbow trout in lentic ecosystems.

A recent study in a small lentic ecosystem [9] found that the rainbow trout had an average relative condition factor of 72.5% (± 2.1%), while the cutthroat trout had an average of 101.0% (± 4.9%). This result suggests that rainbow trout might not be outcompeting cutthroat trout in lentic ecosystems. However, that study had relatively few samples from a small body of water. The purpose of the present study was to more thoroughly test the hypothesis that rainbow trout do not outcompete cutthroat trout in a lentic ecosystem. Three taxa of fish were selected for this study: Snake River cutthroat trout (*Onchorynchus clarkii bouvieri*), rainbow trout (*Onchorynchus mykiss*) and the hybrid cutbow trout (*Onchorynchus mykiss x clarkii*). This study reports that both the original data from Eleven Mile Reservoir as well as milti-year data compiled by the Colorado Division of Parks and Wildlife (CDPW) show that rainbow trout and their hybrids consistently have lower relative condition factors not only in the primary study location (Eleven Mile Reservoir, Colorado), but also in three other Colorado reservoirs (Twin Lakes Reservoir, Turquoise Reservoir, and Antero Reservoir) where cutthroat trout and rainbow trout are found together.

## 2. Methods

Creel surveys were performed at Eleven Mile State Park, Colorado, from May through October, 2012. Eleven Mile Reservoir is a mountain reservoir with a surface area of 13.78 km$^2$, a length of 8.9 km, and a depth that varies from about 7m to about 20m. Length and weight data were recorded for the three taxa. The relative condition factor was computed for each fish in a spreadsheet program. The condition factors were then plotted by taxon and month for comparison.

Permission for creel surveys was obtained from the Colorado Division of Parks and Wildlife (CDPW). Anglers were asked for permission to weigh and measure their fish on a voluntary basis. Total length and fork length were measured to the nearest 3.2 mm (1/8 inch) with a steel tape measure. Depending on whether fish could be brought to the measuring table, weight was measured to the nearest 2.27 g on a market scale (Berkel DX342 Digital Scale) or to the nearest 10 g on a hanging scale (Berkley FS-15). Calibration data was taken for each scale for each creel survey day by measuring readings for calibration weights. Data was recorded by hand in a field notebook and later transferred into a spreadsheet program. Relative condition factor was computed for each fish.

Relative condition factor (Kn) is often used in fisheries and is based on weight and length of the animal. It is assumed that a heavier fish (higher relative condition factor) is a healthier fish, because extra weight means extra energy reserves. It is a reasonable observation that plumper fish are likely receiving a higher share of the available forage, or possibly receiving comparable forage with lower energy expenditure. A "normal" relative condition factor is 1.00, or 100%. The relative condition factor, Kn, for a specific fish is its actual weight divided by its expected weight for its length, based on a reference weight-length equation. A normal condition factor is 1.00, a higher value means a fish is heavier than expected, and a lower value means a fish is lighter than expected. The relationships for expected weights in Colorado were obtained from the CDPW (Table 1).

For each taxon the average relative condition factor was computed for each month. The uncertainty for each value was also computed as the standard error of the mean. The average relative condition factors were plotted for each month and each taxon from May through October 2012. Relative condition factors were also computed for cutbow trout and cutthroat trout data measured in 2011 and plotted for the available months.

In addition to analysis of original data, the relative condition factors were also computed from survey data from the CDPW from 2003 to 2010. The CDPW survey data for Eleven Mile Reservoir was obtained by their biologist each year using nets. Relative condition factors were computed and plotted in a similar manner for the CDPW data as for the present study.

For further comparison with historical weight-length expectations, data from Carlander [10] for rainbow trout and cutthroat trout were used. Methods described in Courtney et al. [9] were used to generate mean weight-length curves, as well as 25$^{th}$ and 75$^{th}$ percentile curves for each taxon. Minimum and maximum curves were also generated from the Carlander data using the fish with the minimum and maximum weights for each 25 mm length interval. To reduce the possibility of length related bias, the Carlander data from 150 mm to 600 mm total length was used to produce the comparison curves.

Table 1: Expected weight relationships used to compute relative condition factors, with weights in g and total lengths in mm.

| Taxon | Expected Weight* |
|---|---|
| Cutthroat Trout | $W(L) = 10^{-5.5304}L^{3.2101}$ |
| Cutbow Trout | $W(L) = 10^{-5.5134}L^{3.2178}$ |
| Rainbow Trout | $W(L) = 10^{-4.898}L^{2.99}$ |

*Private communication from Colorado Division of Parks and Wildlife

Figure 1: Relative condition factors (and uncertainties) by month in Eleven Mile Reservoir, Colorado in 2012 (A) and 2011 (B). Uncertainties are the standard error of the mean.

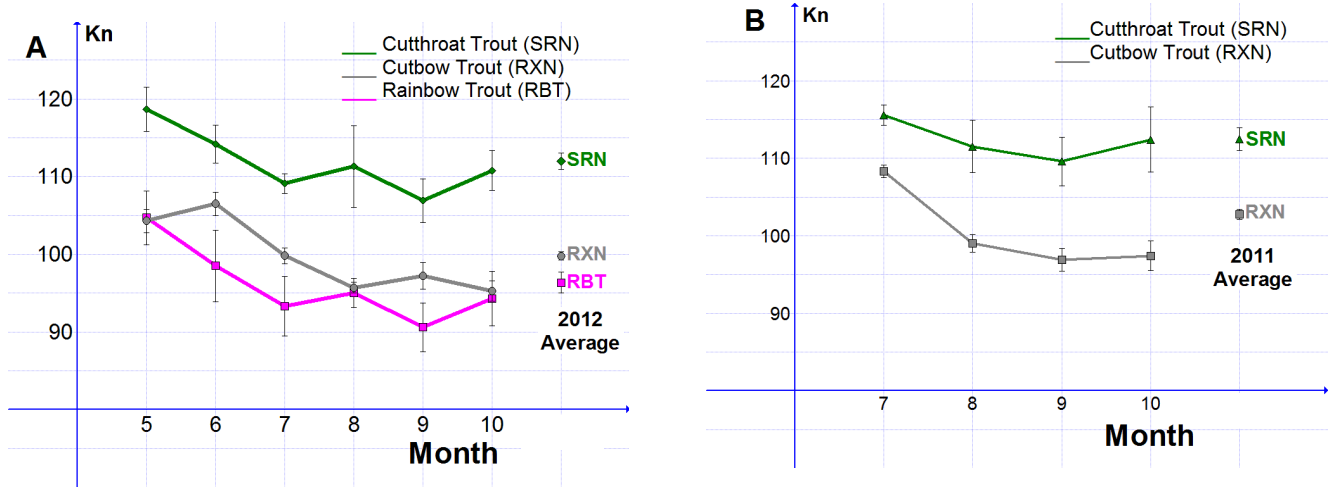

### 3. Results

The cutthroat trout were plump: the mean relative condition factor of the cutthroat trout measured over the six-month study period was 112.0% (± 1.0%). The cutbow trout hybrid were close to the expected weights with a mean relative condition factor of 99.8% (± 0.6%). The rainbow trout were thinner with a mean relative condition factor of 96.4% (± 1.4%). In May, 2012, all three taxa measured had average relative condition factors greater than 100%. Plotting the mean relative condition factors by month (Figure 1A) shows that the relative condition factor of each taxon generally declined during the study.

    Data collected from July through October, 2011, for cutthroat trout and cutbow hybrid trout show a similar pattern. The cutthroat trout had higher relative condition index (average 112.4% ± 1.5%) than the cutbow hybrid trout (102.7% ± 0.6%). Plotting the mean relative condition factors by month (Figure 1B) shows that the relative condition factor of each taxon generally declined over the study period, but the overall average condition factors were greater than 100%.

    Comparing mean relative condition factors computed using CDPW data for earlier years (Figure 2) shows that the native cutthroat trout (SRN) consistently had a higher relative condition factor than the cutbow trout hybrid or the rainbow trout. Weight and length data were usually collected once per year in June.

    Plotting the 2012 data relative to percentile curves generated from Carlander [10] (Figures 3-5) also shows the same trend of cutthroat trout, cutbow trout and rainbow trout having healthy weights for their lengths. The best fit weight-length curve for the cutthroat trout is close to the maximum Carlander curve (Figure 3). Several specimens of cutthroat trout weighed and measured in this study had weights greater than the heaviest fish recorded by Carlander for a given length.

**Figure 2:** Relative condition factors (and uncertainties) by year for cutthroat trout, cutbow trout and rainbow trout in Eleven Mile Reservoir, Colorado. Uncertainties are the standard error of the mean.

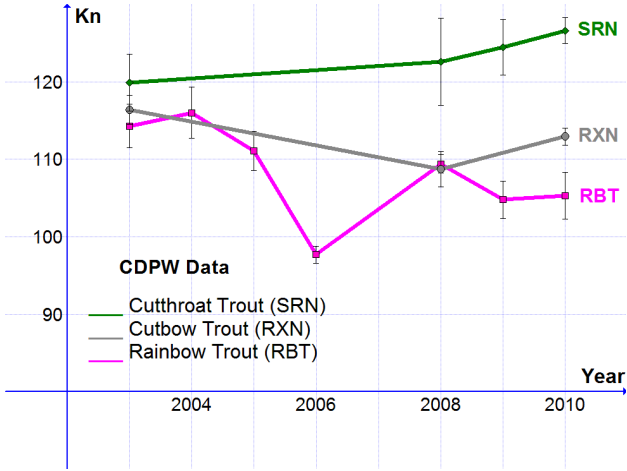

**Figure 3:** Weight-length data for cutthroat trout in Eleven Mile Reservoir, 2012. Best fit curve for weight in grams based on length in millimeters is shown as a dashed line: $W = 1.2479 \times 10^{-5} L^{2.9823}$; $R^2 = 0.9152$. Min, max and percentile curves based on data for cutthroat trout in Carlander [10] are shown for comparison.

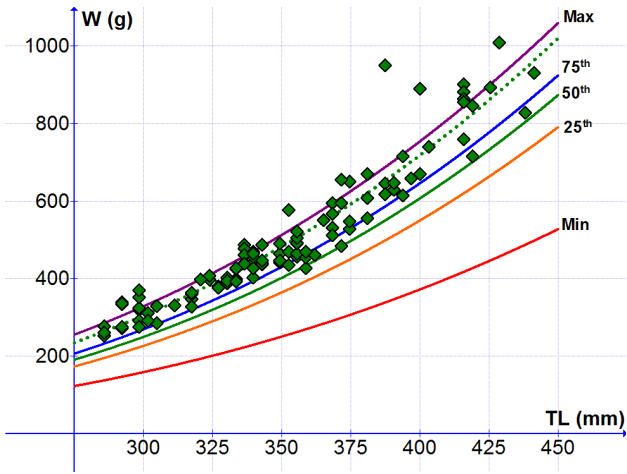

The best fit weight-length curve for the cutbow trout hybrid is near the 50th percentile Carlander curve (for rainbow trout, Figure 4). The weights of the cutbow trout measured in 2012 were clustered between the 25th and 75th percentile curves, with no fish close to the maximum or minimum weights recorded by Carlander.

The best fit weight-length curve for the rainbow trout is just under the 75th percentile Carlander curve (Figure 5). All rainbow trout measured in 2012 were heavier than the 25th percentile Carlander curve, and several were well above the 75th percentile Carlander curve, but none were heavier than the maximum Carlander curve.

**Figure 4**: Weight-length data for cutbow trout in Eleven Mile Reservoir, 2012. Best fit curve for weight in grams based on length in millimeters is shown as a dashed line: $W = 2.7706 \times 10^{-5} L^{2.8451}$; $R^2 = 0.9401$. Min, max, and percentile curves based on data for rainbow trout in Carlander [10] are shown for comparison, since data for the hybrid were not available.

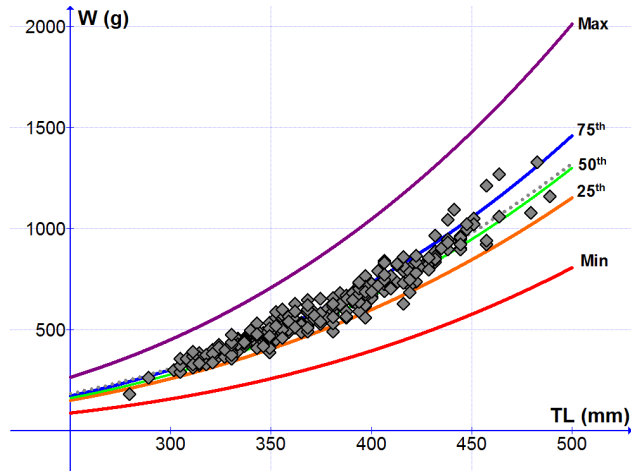

**Figure 5**: Weight-length data for rainbow trout in Eleven Mile Reservoir, 2012. Best fit curve for weight in grams based on length in millimeters is shown as a dashed line: $W = 1.1275 \times 10^{-5} L^{3.0021}$; $R^2 = 0.9205$. Min, max, and percentile curves based on data for rainbow trout in Carlander [10] are shown for comparison.

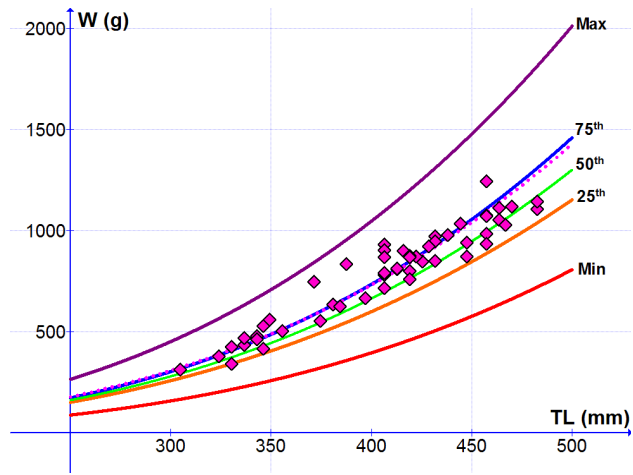

**4. Discussion**

Much has been written about introduced rainbow trout interbreeding and outcompeting native cutthroat trout; though the published data is limited to lotic ecosystems (rivers and streams). Results of a small study in Colorado [9] suggested that rainbow trout might not outcompete cutthroat trout in a lentic ecosystem. In the present study, several independent approaches were taken to more thoroughly investigate this hypothesis.

Data from creel surveys performed from May through October, 2012, on cutthroat trout, cutbow trrout and rainbow trout supports the hypothesis that rainbow trout are not strong food competitors with cutthroat trout in lentic ecosystems. On the contrary, the data suggests that cutthroat trout are outcompeting both the cutbow trout hybrid and rainbow trout in this lentic ecosystem, though all three taxa had healthy relative condition factors. Historical data from creel surveys peformed from July through October, 2011, on cutthroat trout and cutbow hybrid trout showed similar results.

Data provided by the CDPW and collected from 2003 to 2010 for the same body of water were used to compute the average relative condition index for each taxon and year. Most years, data were collected in June. Collection methods varied from year to year and included gill netting and trap netting. The collection method can affect assessment of fish populations because one may be more likely to catch fish of a certain size with a specific method. However, the use of relative condition factor to assess overall health of fish takes size into consideration and so is not as sensitive to the collection method. For these eight years, the cutthroat trout had a higher average relative condition factor than the cutbow trout or rainbow trout. Data collected for this study in 2012 were compared to weight-length relationships published by Carlander [10]. Compared to weight-length equations published by Carlander, the cutthroat trout in particular were above the Carlander 50th percentile curve and included several specimens that were heavier than the heaviest specimens recorded by Carlander for a given length.

A key question is whether the propensity for cutthroat trout to be plumper than rainbow trout and their hybrids when they are in the same lentic systems is limited to Eleven Mile Reservoir and Dead Man's Lake [9] or whether there is something peculiar to these systems giving cutthroat trout a competitive advantage that is not present in most lentic systems. To shed light on this question, data was requested from CDPW for three additional Colorado reservoirs where cutthroat and rainbow trout are found in sympatry. Year to year condition factors were computed as described above and are shown in Figures 6, 7, and 8. The general trend observed in Eleven Mile Reservoir is also seen in data from Antero Reservoir, Twin Lakes Reservoir, and Turquoise Reservoir.

The notable exception is 2009 in Twin Lakes Reservoir and Turquoise Reservoir, as shown in Figures 7 and 8. These two reservoirs are unique among eastern slope reservoirs in that they are fed by snowfall from the western slope of the Rocky Mountains which is diverted from the Fryingpan watershed on the western slope via a series of tunnels to the Arkansas watershed on the eastern slope. As it happens, 2009 was a record snowfall year for the upper Fryingpan basin, and this led to record inflows into these two reservoirs from March to May 2009. The inflow into the reservoirs in this time period represented over 100% of the storage volume and may have had the effect of preventing stratification and production of the usual food web. Under the unusually large flow conditions these two reservoirs may have been something of a hybrid between lotic and lentic ecosystems resulting in closer to even competition between cutthroat and rainbow trout.

Together with the data from Eleven Mile Reservoir and Dead Man's Lake, these three additional systems demonstrate the tendency for cutthroat trout to have higher body condition in a variety of systems including high elevation smaller bodies of water that are sympatric with lake trout (Turquoise Reservoir and Twin Lakes Reservoir), mid-elevation larger reservoirs that are sympatric with northern pike (Eleven Mile Reservoir and Antero Reservoir), and a small lower elevation lake where there is not a fish predator higher on the food web than cutthroat trout and rainbow trout (Dead Man's Lake). Cutthroat trout seem to be more plump in less productive systems such as Dead Man's Lake, Turquoise Reservoir, and Twin Lakes Reservoir where rainbow trout tend to have average condition factors under 90%, as well as in highly productive reservoirs (Antero Reservoir, Eleven Mile Reservoir) where cutthroat trout, rainbow trout, and their hybrids all average over 100% in relative condition factor. The tendency for cutthroat trout to be significantly plumper than rainbow trout in sympatric lentic systems does not seem to depend on details of the lentic system, except for the unusual very high flow conditions of 2009 in Turquoise Reservoir and Twin Lakes Reservoir.

A study in British Columbia, Canada [11] compared food, size and growth between rainbow trout and cutthroat trout from lake populations. They studied whether being sympatric or allopatric affected the relative growth and food habits of rainbow trout and cutthroat trout. In sympatric populations, the cutthroat trout were larger at specific ages compared to rainbow trout. The opposite was found for allopatric populations. This result further supports that the competitive relationship between cutthroat trout and rainbow trout could be different in lentic and lotic ecosystems.

In conclusion, these three lines of inquiry, along with the results of Nilsson and Northcote [11], suggest that rainbow trout do not outcompete cutthroat trout in lentic ecosystems, which is different from what has been observed for lotic ecosystems. Besides being important to understand scientifically, there are potential implications of this result. For example, it suggests the possibility of incorporating lentic areas into lotic ecosystems for the purpose of improving cutthroat trout condition or balancing the competition between the taxa in a geographic area.

**Figure 6:** Relative condition factors (and uncertainties) by year for cutthroat trout, cutbow trout and rainbow trout in Antero Reservoir, Colorado.

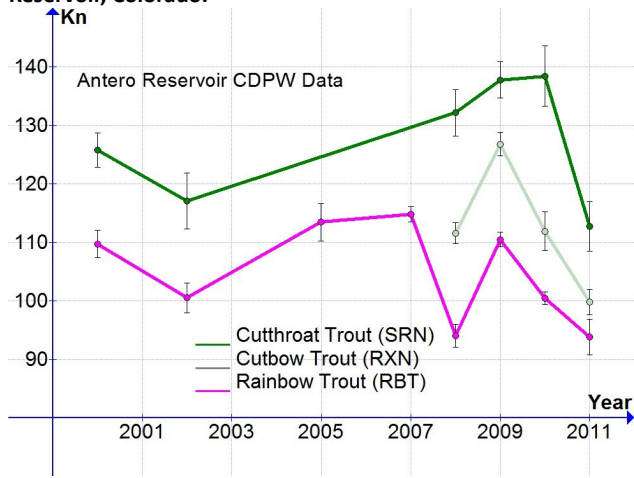

**Figure 7:** Relative condition factors (and uncertainties) by year for cutthroat trout and rainbow trout in Turquoise Reservoir, Colorado.

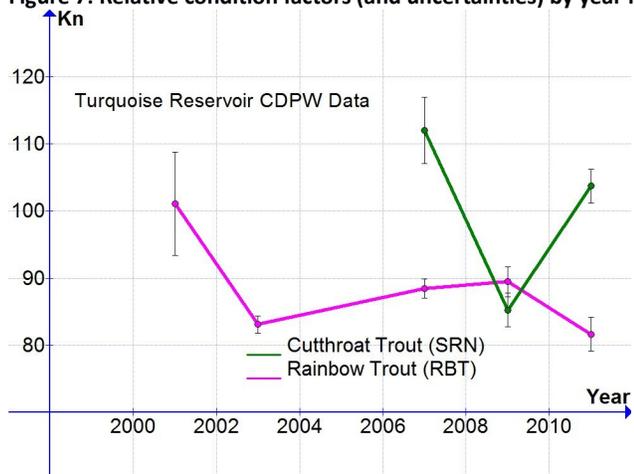

**Figure 8:** Relative condition factors (and uncertainties) by year for cutthroat trout, cutbow trout, and rainbow trout in Twin Lakes Reservoir, Colorado.

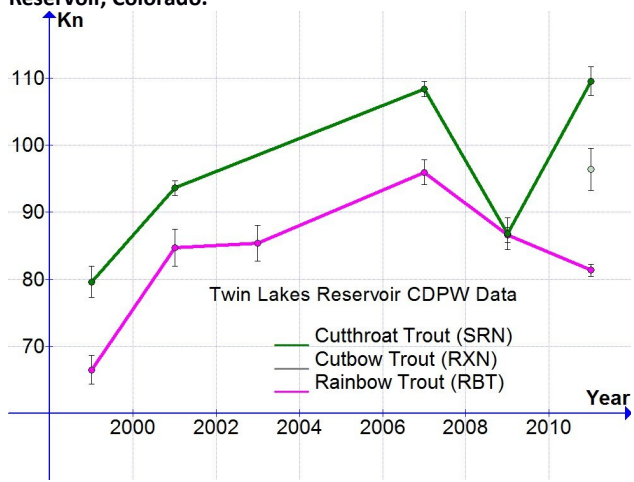

**Competing Interests**
The authors declare that they have no competing interests.

**Acknowledgments**
The authors appreciate the cooperation of the management and staff at Eleven Mile State Park, especially Kevin Tobey, who granted permission for the creel surveys and the staff at the boat inspection station who are always helpful, friendly, and efficient. Larry at the 11 Mile Marina was also helpful. The authors are thankful to all the anglers and park guests who allowed us to weigh and measure their fish. Harry Vermillion (CDPW) provided the data attributed to the Colorado Department of Parks and Wildlife.

**Author Contributions**
JMC conceived and designed the experiments. JMC performed the experiments with assistance from MWC and ACC. JMC analyzed the original experimental data. JMC and MWC analyzed the CDPW data. ACC contributed materials/analysis tools. ACC and JMC wrote the early drafts of the paper. MWC edited the paper.